%% file: main.tex
\documentclass{article}

\usepackage{PRIMEarxiv}

\usepackage[utf8]{inputenc} 
\usepackage[T1]{fontenc}    
\usepackage{listings}
\usepackage{hyperref}       
\usepackage{url}            
\usepackage{booktabs}       
\usepackage{amsfonts}       
\usepackage{nicefrac}       
\usepackage{microtype}      
\usepackage{lipsum}
\usepackage{fancyhdr}       
\usepackage{graphicx}       
\graphicspath{{media/}}     

\title{Type Checking Project Haystack Grids \\
using JSON Schema and Pydantic
}

\author{
  Thomas Hirsch \\
  Affiliation \\
  TU Wien \\
  Vienna, Austria\\
  \texttt{thomas.hirsch@tuwien.ac.at} \\
   \And
  Samina Kadkhoda Masoumali \\
  Affiliation \\
  TU Wien \\
  Vienna, Austria\\
  \texttt{samina.masoumali@tuwien.ac.at} \\
   \And
  Gerald Schweiger \\
  Affiliation \\
  TU Wien \\
  Vienna, Austria\\
  \texttt{gerald.schweiger@tuwien.ac.at} \\
}

\begin{document}
\maketitle

\begin{abstract}
\input{sections/abstract}
\end{abstract}


\input{sections/intro}
\input{sections/relatedWork}
\input{sections/problem}
\input{sections/implementation}

\input{sections/conclusion}

\section*{Acknowledgments}
The research leading to these results was conducted within the
project Self2B (project number 920143), funded by the Austrian
Research Promotion Agency (FFG).

\bibliographystyle{unsrturl}
\bibliography{building_ontologies}

\end{document}

%% file: sections/abstract.tex
Ontologies enable scalable energy services in buildings by supporting interoperability and automation. 
Project Haystack is a building ontology that is widely adopted due to its flexible, tag-based semantic model, openness, and extensibility, but suffers from ambiguous tag usage and limited automated validation.
Although Project Haystack is formally open, its reliance on custom file formats and domain-specific languages that originate from the Haxall ecosystem creates a de facto barrier to integration. In this paper, we address these limitations by introducing a Python-based toolchain for Haystack. We present (i) a parser for Haystack definition files (Trio file format), and (ii) a code generator that derives Pydantic models and JSON Schema definitions from these parsed specifications. The resulting models enable static type checking and enable structural validation of Haystack grids within Python, as well as schema-based validation of JSON representations outside the Python ecosystem. All tools, generated models, and schemas are released publicly under an open-source license, with the goal of strengthening the Haystack ecosystem and opening a practical pathway beyond its current technical boundaries.


%% file: sections/intro.tex
\section{Introduction}

Ontologies are a key enabler for scalable energy services in buildings \cite{alfalouji2022iot}. They provide a semantic layer that supports interoperability and automation across projects, tools, and stakeholders. 
Ontologies have been investigated and compared in terms of coverage, expressiveness, interoperability, and transferability \cite{ quinnCaseStudyComparing2021,paulsonComparisonBrickSchema2021, qiangSystematicComparisonEvaluation2023, luoOverviewDataTools2021,  pengBuildingOntologies45GDHC2025,quinnComparisonBrickProject2022, lygerakisKnowledgeGraphsOntologies2022, pritoniMetadataSchemasOntologies2021}.

Project Haystack Version 4\footnote{Versions 3.9.x were the development/preliminary stages of Version 4 and were already in widespread use before Version 4.0 was released} (further referred to only as Project Haystack in this work) is a promising ontology that has achieved industrial adoption due to its intuitive tagging-based semantic model, openness, flexibility, and extensibility \cite{quinnCaseStudyComparing2021}.
It defines a semantic tagging model for building data\cite{HomeProjectHaystack, quinnCaseStudyComparing2021}. To use this model in operational systems, a Haystack-compliant server implementation such as Haxall or the commercial SkySpark platform is required to store, manage, and expose the tagged data via standardized APIs \cite{HomeProjectHaystack, Architecture}. Client libraries and formats, such as the Phable Python client or the JSON-based Hayson representation, are then used to query, exchange, and integrate Haystack data with external applications.

Project Haystack's flexibility, included query/filter language that is easy to use, organization structure and steering board comprised of established companies in the field of building technologies and control \cite{ProjectHaystack}, as well as the availability of reference software implementations including a full ontology server including a database system explains its acceptance and spread.

Despite these strengths, Haystack exhibits several limitations, some related to the ontology itself, some related to the ways in which the ontology is defined and specified, and some related to the software ecosystem around Project Haystack.
This results in ambiguities and inconsistencies in tag usage, and the lack of systematic, automated mechanisms to validate models against the standard, and hardships to implement and add such validation mechanisms~\cite{pengBuildingOntologies45GDHC2025, luoOverviewDataTools2021, bergmannSemanticInteroperabilityEnable2020, fierroHouseSticksFormalizing2019, quinnCaseStudyComparing2021, fierroFormalizingTagBasedMetadata2020, quinnComparisonBrickProject2022, yefiMetamEnThObjectOrientedMetamodel2024, bhattacharyaShortPaperAnalyzing2015}.

Although Project Haystack is formally open, its reliance on custom file formats and domain-specific languages (e.g., Trio, Zinc, Xeto; Fantom, Axon) that originate from the Haxall ecosystem creates a de facto barrier to integration.
Haxall is the reference implementation of a Haystack server \cite{Haxall, HaxallInitiativeSkyFoundry}. 
The company behind Haxall is Skyfoundry, offering their proprietary Skyspark architecture that builds upon Haxall and offers scalability, sensor data recording, and analytics services.

To illustriate how intertwined Project Haystack is with the Haxall ecosystem:
The source code of the schemantic tagging models core definitions is provided in the form of  Trio files, "a semi-subset of YAML" \cite{TrioProjectHaystack}.
Building the actual ontology / applicable definitions from these sources requires the \verb|DefCompiler|.
The reference Parser for Project Haystacks custom Trio file format, as well as the \verb|DefCompiler| - are part of Haxall.
Haxall and all of its components are implemented in the Fantom \cite{HomeFantom} programming language.

Beyond forum and mailing list posts there is no discussion on the software engineering perspective 
Interfacing Haystack with mainstream programming languages and data pipelines remains cumbersome, and available parsers and compilers for Haystack definition formats are largely confined to the Fantom ecosystem.

Alongside missing validators, there is a lack of software supporting creation and construction of schemantic models of buildings using the Project Haystack ontology.
This lack of validation together with the flexible tagging approach creates an interoperability and transferability problem.
While the Xeto specification language is intended to fill this role in the future, development is still ongoing (part of Haystack 5), and Project Haystack specifications in the form of Xeto specifications is not yet complete.
Once finished Xeto may be used to validate Haystack models; however, it is again implemented in the Fantom programming language.

\subsection{Main contribution}
In order to increase acceptance of the Project Haystack ontology, we argue that it has to become more open in a technical sense, by providing parsers, tools, validators in general purpose programming languages other than Fantom - beyond interfaces and bindings to Fantom implemented tools.
Endeavors like J2inn's Haystack definitions to typescript interfaces creator, the phable python client for Haxall servers, the JSON schema definitions of Haystack definitions, and a multitude of other smaller projects in (java, c++, etc) supports our thought that there is a real need for using / processing Haystack models in programming languages and frameworks outside the Fantom/Haxall ecosystem.

In this paper, we address these limitations by introducing a Python-based toolchain for Haystack. We present (i) a parser for Haystack definition files written in Trio, and (ii) a code generator that derives Pydantic \cite{WelcomePydanticPydantica} models and JSON Schema definitions from these parsed specifications. The resulting models enable static type checking and enable structural validation of Haystack grids within Python, as well as schema-based validation of JSON representations outside the Python ecosystem.

Applying this approach, we identify inconsistencies and missing definitions in selected Haystack releases, demonstrating the practical value of automated validation.

By enabling type checking and partial validation of Haystack models using widely adopted technologies, our work improves interoperability, lowers the entry barrier for new users, and facilitates integration with modern software stacks. All tools, generated models, and schemas are released publicly under an open-source license, with the goal of strengthening the Haystack ecosystem and opening a practical pathway beyond its current technical boundaries.
All code and data, including the Pydantic models and JSON schemas are made publicly available under an open source license on github.\footnote{
\href{https://github.com/tuw-isab/phaystackschema}
{
\url{https://github.com/tuw-isab/phaystackschema}
}
}







%% file: sections/relatedWork.tex
\section{Related Work}

Quinn et al. \cite{quinnCaseStudyComparing2021} conduct a qualitative comparison of the Brick Schema and Project Haystack ontologies. They observe comparable completeness and expressiveness for both, though Haystack under-performs across six key ontology attributes: flexibility, portability, readability, extensibility, interoperability, and queryability.

Bergmann et al. \cite{bergmannSemanticInteroperabilityEnable2020} proposes a pathway to advance semantic interoperability for smart buildings. They discuss and compare various building metadata schemas, highlighting brick and project Haystack as the predominant operational metadata schemas, while criticizing Haystack for unverifiable correctness and extensive customizability, eroding confidence in its implementations.

Fierro et al. \cite{fierroHouseSticksFormalizing2019} perform a qualitative analysis of Project Haystack and identifies interpretability and consistency issues caused by its lack of formal semantics. They introduce Brick+, a formally defined ontology that enables reliable inference of Brick models from Haystack data, demonstrated on five models.

Qiang et al. \cite{qiangSystematicComparisonEvaluation2023} perform an in-depth systematic evaluation and comparison four building ontologies (Brick Schema, RealEstateCore, Project Haystack, and Digital Buildings) based on axiomatic design and assertions in a use case. Further, they investigate the compatibility between ontologies and propose paths towards alignments and translations between these different ontologies.

Luo et al. \cite{luoOverviewDataTools2021} reviews 24 building data tools across the building lifecycle, categorizing them as data dictionaries/terminologies, ontologies/schemas, and platforms. They discuss brick and project Haystack within their ontology category. 

Pen et al. \cite{pengBuildingOntologies45GDHC2025}  investigate the modeling capabilities of six major building ontologies (Brick, RealEstateCore, Project Haystack, SAREF, Flow Systems Ontology, and ASHRAE 223P) for 4th–5th generation district heating and cooling (45GDHC) systems.
They show that while existing ontologies handle basic building and sensor data well, they fall short in representing district-level energy exchanges and prosumer relationships.

Yefi et al. \cite{yefiMetamEnThObjectOrientedMetamodel2024} investigate Brick and Project Haystack ontologies capabilities to model building systems beyond hvac systems. They identify seven key issues with the existing ontologies, amongst them lack of validation mechanisms, and semantic issues relevant to this work, and propose their own object oriented meta model.

A recurring theme in the related work on building ontologies is the project Haystack ontology being more widely adopted, performing en-par with other building ontologies when it comes to completeness and expressiveness, while suffering from its lack of formality, machine readability, and verifiability \cite{pengBuildingOntologies45GDHC2025, luoOverviewDataTools2021, bergmannSemanticInteroperabilityEnable2020, fierroHouseSticksFormalizing2019, quinnCaseStudyComparing2021, fierroFormalizingTagBasedMetadata2020, quinnComparisonBrickProject2022, yefiMetamEnThObjectOrientedMetamodel2024, bhattacharyaShortPaperAnalyzing2015}.

Several of the works mentioned above base their investigations on Project Haystack Version 3, explicitly noting that Version 4 was under development and expected to address some of the identified limitations \cite{luoOverviewDataTools2021, bergmannSemanticInteroperabilityEnable2020}.

This work, is based in Project Haystack Version~4. In the following section, we demonstrate that many of the issues observed in earlier versions persist, although possibly to a lesser extent.





%% file: sections/problem.tex
\section{Design Challenges}
This section presents some of the design challenges we faced when trying to implement basic means of automated validation for haystack models.
Some are general issues with Project Haystack ontology definitions, others arise from our target to achieve this without relying on the Haxall/Fantom ecosystem.

\subsection{Integration Challenges}
\textbf{Trio}
The Trio file format \cite{TrioProjectHaystack} is the primary format for project haystack definitions \cite{ProjectHaystackHaystackdefs2025}.
While Trio was based on YAML, it is not valid YAML, and YAML parsers can not be used to read Trio files.
There is no open source Python implementation of a Trio parser available.



\textbf{Build process}
The process of building the project haystack ontology is called normalization \cite{NormalizationProjectHaystack}.
The reference implementation of normalization process is the DefCompiler, implemented in the Fantom programming language and contained in the Haxall project \cite{HaxallHaxall2026}.
This build process is rather complex (e.g. the \verb|defx| system \cite{NormalizationProjectHaystack} for late binding).

\textbf{Broken versions}
The tagged release of project haystack version 3.9.8 is invalid due to missing type definitions. There seems to be a lack of testing and validation during and after the normalization process for versions before 3.9.10. The validation process described in \cite{NormalizationProjectHaystack} is minimal and rather simplistic, and does not capture many of the ontological challenges described below.

\textbf{Serialization of Haystack models}
Haystack models of buildings and technical equipment within them can be serialized and stored in JSON format \cite{JsonProjectHaystack}, aside from the Project Haystack specific formats (Trio and Zinc).

J2inn \cite{J2innHayson2025} provide a publicly available JSON Schema to validate this so called "Hayson" schema that is of substantial value for software developers working outside the Haxall/Fantom ecosystem.
However, this schema validates only the general serialization format, it does not validate the contained models adherence to the Project Haystack ontological definitions.

\subsection{Ontological Challenges}

\textbf{Definitions not machine readable}
Some definitions are not provided in machine readable form, with information provided only in prose text form in the Project Haystacks documentation, not being reflected in the corresponding Trio files.

Notable examples can be found in the point definitions:

"All points must define exactly one of the following pointFunction marker tags"\cite{PointsProjectHaystack}
Which is not reflected in the Trio definitions of the type point or \verb|pointFunction|.
The same pattern reoccurs with the \verb|kind| parameter of a point.
Even the currently under development Xeto definitions of points do not reflect the documentation. While \verb|kind| is indeed mandatory in the Xeto definitions, the \verb|pointFunction| is not - contradicting the prose text documentation.

Being only one of many examples, the arising situation requires a developer of a haystack model for a building or system to have in-depth knowledge of the haystack standard and its documentation, and will inevitably lead to a broader spectrum of what different applicants of the haystack ontology deem a valid haystack model.

Further, such lapses in lack of machine readable ontology definition lead to inconsistent and even contradictory ontology definitions.
For example points connected to \verb|weatherStation| \cite{WeatherProjectHaystack} contradict the definition of \verb|point| \cite{PointsProjectHaystack} and its mandatory fields.

% ...


Missing, or incomplete machine readable definitions prevent automated validation and checking of the ontological definitions and models based on this ontology.
Previous investigations have already highlighted this issue \cite{bergmannSemanticInteroperabilityEnable2020, fierroHouseSticksFormalizing2019} in previous haystack versions, however, this problem still persists in Project Haystack 4, and in the current state of Project Haystack 5 (Xeto).

\textbf{Loose definitions}
Some haystack types are defined very loosely, complicating type checking and validation.
Notable example of this is the \verb|enum| type, which can be applied to \verb|def| as well as \verb|point|, and can contain \verb|dict|, \verb|str|, Xeto \verb|ref|, as values. Furthermore, if the value of an \verb|enum| is of type string, it can be a comma separated, newline separated, or markdown formatted list according to its prose text documentation \cite{EnumProjectHaystack}.


While versions before 3.9.15 defined enum with \verb|is:[^str,^dict]|, newer versions simply define it as its own type using \verb|val| for which one must read the prose text documentation to determine the allowed data types.
Finally, depending on whether \verb|enum| is applied to a \verb|def| or a \verb|point| results in very different handling when modeling a building grid despite being of the same type underneath.

\textbf{Ambiguous Cardinality}
The Trio format and the defs are not expressive enough to define cardinality of relationships/references.
There is no way to distinguish one-to-one and one-to-many relationships in Project Haystack definitions.
For example:
The project haystack documentation states that the basic containment relations (\verb|siteRef|, \verb|equipRef|, or \verb|spaceRef|) should always be unary \cite{FiltersProjectHaystack}, while \verb|systemRef| is explicitly allowed to contain a list of refs \cite{SystemsProjectHaystack}.
The Trio source definitions of \verb|systemRef| and \verb|equipRef| are both defined as \verb|is:[^ref]|.
In contrast, \verb|siteRef| is defined as \verb|is: ^ref| definition.
While at first glance, the \verb|equipRef| seems to be incorrect, allowing for multiple ref despite the documentation allowing only unary ref, the issue is more complex:
During normalization all \verb|is: ^ref| are transformed to \verb|is:[^ref]|.
In the normalized ontology, \verb|is:[^ref]| is to be interpreted as being either list of ref, or ref.
Finally, \verb|[]| are not only used to imply cardinality, but also to denote a def having multiple supertypes (e.g. battery: \verb|is: [^equip, ^elec-output]|)




Haystack 5 introduces Xeto to strengthen the ontology. For example, the situation described above can now be solved with the Xeto type \verb|MultiRef|. However, Xeto is still in development, comes with yet another new custom .xeto file format, and corresponding Xeto programming language that itself is implemented in Fantom.

%% file: sections/implementation.tex
\section{Implementation}
In order to create type checking information in the form of JSON and Pydantic schemas the Project Haystack definitions have to be read.
From version 3.9.10 on, normalized definitions are included in the Project Haystack releases. Normalized definitions are available in Trio, Zinc, JSON, JSONld and ttl formats.
For versions before 3.9.10 only the Trio source files are available.
We therefore implement two paths for Pydantic and JSON Schema creation:
\begin{itemize}
    \item A rudimentary re-implementation of the normalization process described in Project Haystack.
    \item A transformation of the normalized definitions obtained in JSON format.
\end{itemize}

The normalization process as defined in Project Haystack
\\documentation \cite{NormalizationProjectHaystack}:
\begin{enumerate}
\item Scan: traverse input libs to find Trio files
\item Parse: parse each Trio file to discover def and defx dicts
\item Resolve: ensure every symbol tag maps to a def
\item Taxonify: compute taxonomy tree of supertype/subtypes
\item Defx: add defx tags to each declared def
\item Normalize Tags: normalize def tags
\item Inherit: recursively apply supertype tags into subtypes
\item Validation: perform additional sanity checks
\item Generate: output Namespace
\end{enumerate}

Using our implementations we create schemas from both sources (if available) and compare them to ensure that our normalization process and Trio parser implementation are aligned with the normalization process implemented in \verb|DefCompiler| in the Haxall repository.

\textbf{Step 1(a) Trio source}
We implement a Trio parser using python lark \cite{LarkparserLark2026}.
Using this parser we read all Trio files in the Haystack definitions.
We only consider tag definitions that are generally applicable or applicable to rows in Haystack model.
We apply the def extension (\verb|defx|) to perform the late binding step, updating the parsed definitions.

\textbf{Step 1(b) Normalized Json source}
The normalized Haystack ontology is loaded from a release (3.9.10 and above).

\textbf{Step 2 Resolve inheritance}
We calculate the inheritance tree for all defs to deduce the base datatype for each tag.
The intermediate results are stored in a JSON file.

\textbf{Step 3 Filter}
The resulting definitions are filtered based on \verb|tagOn| field, origin library and namespace, and the inheritance tree of each definition.
Only definitions that are applicable to rows in a Haystack model are kept.

\textbf{Step 4 Schema creation}
We implement Pydantic \cite{WelcomePydanticPydantica} schemas for each base type and created a mapping from Project Haystack types to those schemas.
This mapping is applied to the inheritance trees of the filtered defs to create a Pydantic schema of a row in a Haystack model.
This row schema now contains all tags/definitions that are allowed to occur in a Haystack models row together with the correct datatype.
This row model is wrapped in another schema representing a Haystack grid.

An excerpt from the resulting Pydantic schema:

\begin{lstlisting}

class HsGridRow(BaseModel):
  id: HsRef
  dis: str = Field(..., title='Dis')
  ahu: Optional[HsMarker] = None
  equip: Optional[HsMarker] = None
  siteRef: Optional[Union[HsRef, HsRefList]] = Field(None, title='Siteref')
  geoAddr: Optional[str] = Field(None)
  geoCity: Optional[str] = Field(None)
  site: Optional[HsMarker] = None
  area: Optional[Union[HsNumber, float]] = Field(None, title='Area')
  ...
\end{lstlisting}

This Pydantic schema can now be used to type check a Haystack model in the form or a JSON formatted Haystack grid.
An exemplary application:
\begin{lstlisting}
with open(path) as fd:
    grid = json.load(fd)
grid = HsGrid.model_validate(grid)
\end{lstlisting}

Pydantic will try to apply the schema to the provided data, if type checking fails, a detailed error message will be provided. If model validation succeeds, a Pydantic model is returned. This Pydantic model behaves like a data-class in python.
For example, a call like \verb|grid.rows[0].ahu| will either return a marker if the first row in the grid has this marker set, or pythons \verb|None| otherwise.

Aside from type checking, using the Pydantic model instead of the python representation of the JSON data structure enables auto-completion capabilities of the IDE, as well as continuous type checking. Setting or accessing an undefined tag will result in an error.
Pydantic models can be serialized back to JSON with ease.

Furthermore, Pydantic allows for serializing the schemas themselves to produce JSON Schema, enabling type checking using JSON Schema, or integration of the schema into many other programming languages and services aside from python.

%% file: sections/conclusion.tex
\section{Conclusion}

Project Haystack has emerged as a practical and well adopted ontology within the building automation industry, primarily due to its intuitive tagging-based semantic model, open design, and high degree of flexibility and extensibility~\cite{quinnCaseStudyComparing2021}. 
Underneath, project haystack consists of a semantic tagging model and filter language for building-related data~\cite{HomeProjectHaystack, quinnCaseStudyComparing2021}

However, despite its strengths, the ecosystem’s close coupling with the Haxall framework and Fantom programming language results in integration challenges. In this work, we address this limitation by introducing a Python-based toolbox that enables users to perform data type checking and partial validation of Project Haystack models independently of the Haxall ecosystem and Fantom programming language. Our approach includes the generation of both JSON Schemas and Pydantic models, allowing developers to seamlessly process, validate, and interact with Haystack data across multiple programming environments.

To achieve this, we implemented a Python parser for the Haystack specific Trio format, re-engineered the normalization and def compilation processes to create an operational Project Haystack ontology, and developed a transformation pipeline producing the corresponding Pydantic and JSON schemas. All code, along with the generated artifacts, is freely available on GitHub\footnote{
\href{https://github.com/tuw-isab/phaystackschema}
{
\url{https://github.com/tuw-isab/phaystackschema}
}
} to promote openness, transparency, and community collaboration.

\textbf{Future work:} We plan to release our implementation as an installable Python package on PyPI, making it easier for developers to access and apply the Haystack schemas in their projects. Further research will focus on the automatic generation of Pydantic validator rules~\cite{ValidatorsPydanticValidation} derived from the machine-readable elements of the Haystack definitions. Additionally, we will develop a dedicated parser for the Xeto format to support automated rule creation and explore the use of SHACL~\cite{ShapesConstraintLanguage2017, SHACLDescriptionLogic, SemanticsValidationRecursive} as a formal mechanism for validating both normalized RDF releases and Project Haystack model instances.

%% file: building_ontologies.bib
@online{Architecture,
  author = {Project Haystack},
  title = {Architecture},
  year = 2026,
  url = {https://haxall.io/doc/docHaxall/Architecture},
  lastaccessed = {2026-01-22},
}

@article{bergmannSemanticInteroperabilityEnable2020,
  title = {Semantic {{Interoperability}} to {{Enable Smart}}, {{Grid-Interactive Efficient Buildings}}},
  author = {Bergmann, Harry and Mosiman, Cory and Saha, Avijit and Haile, Selam and Livingood, William and Bushby, Steve and Fierro, Gabe and Bender, Joel and Poplawski, Michael and Granderson, Jessica and Pritoni, Marco},
  year = 2020,
  month = nov,
  urldate = {2024-11-05},
  langid = {english}
}

@inproceedings{bhattacharyaShortPaperAnalyzing2015,
  title = {Short {{Paper}}: {{Analyzing Metadata Schemas}} for {{Buildings}}: {{The Good}}, the {{Bad}}, and the {{Ugly}}},
  shorttitle = {Short {{Paper}}},
  booktitle = {Proceedings of the 2nd {{ACM International Conference}} on {{Embedded Systems}} for {{Energy-Efficient Built Environments}}},
  author = {Bhattacharya, Arka and Ploennigs, Joern and Culler, David},
  year = 2015,
  month = nov,
  series = {{{BuildSys}} '15},
  pages = {33--34},
  publisher = {Association for Computing Machinery},
  address = {New York, NY, USA},
  doi = {10.1145/2821650.2821669},
  urldate = {2026-01-14},
  isbn = {978-1-4503-3981-0}
}

@online{EnumProjectHaystack,
  author = {Project Haystack},
  year = 2026,
  title = {Enum},
  url = {https://project-haystack.org/doc/lib-ph/enum},
  lastaccessed = {2026-01-29}
}

@article{fierroFormalizingTagBasedMetadata2020,
  title = {Formalizing {{Tag-Based Metadata With}} the {{Brick Ontology}}},
  author = {Fierro, Gabe and Koh, Jason and Nagare, Shreyas and Zang, Xiaolin and Agarwal, Yuvraj and Gupta, Rajesh K. and Culler, David E.},
  year = 2020,
  month = sep,
  journal = {Frontiers in Built Environment},
  volume = {6},
  publisher = {Frontiers},
  issn = {2297-3362},
  doi = {10.3389/fbuil.2020.558034},
  urldate = {2024-11-05},
  langid = {english}
}

@inproceedings{fierroHouseSticksFormalizing2019,
  title = {Beyond a {{House}} of {{Sticks}}: {{Formalizing Metadata Tags}} with {{Brick}}},
  shorttitle = {Beyond a {{House}} of {{Sticks}}},
  booktitle = {Proceedings of the 6th {{ACM International Conference}} on {{Systems}} for {{Energy-Efficient Buildings}}, {{Cities}}, and {{Transportation}}},
  author = {Fierro, Gabe and Koh, Jason and Agarwal, Yuvraj and Gupta, Rajesh K. and Culler, David E.},
  year = 2019,
  month = nov,
  series = {{{BuildSys}} '19},
  pages = {125--134},
  publisher = {Association for Computing Machinery},
  address = {New York, NY, USA},
  doi = {10.1145/3360322.3360862},
  urldate = {2024-11-05},
  isbn = {978-1-4503-7005-9}
}

@online{FiltersProjectHaystack,
  author = {Project Haystack},
  year = 2026,
  title = {Filters},
  url = {https://project-haystack.org/doc/docHaystack/Filters\#refLists},
  lastaccessed = {2026-01-28}
}

@online{Haxall,
  author = {Haxall},
  year = 2026,
  title = {Haxall},
  url = {https://haxall.io/},
  lastaccessed = {2026-01-28}
}

@online{HaxallHaxall2026,
  author = {Haxall},
  year = 2026,
  title = {Haxall/Haxall},
  url = {github.com/haxall/haxall},
  lastaccessed = {2026-01-28}
}

@online{HaxallInitiativeSkyFoundry,
  author = {SkyFoundry},
  year = 2026,
  title = {The {{Haxall Initiative}} - {{SkyFoundry Open Sources Core Software}} to {{Accelerate}} the {{BIoT}}},
  url = {https://www.skyfoundry.com/blog/5924},
  lastaccessed = {2026-01-22}
}

@online{HomeFantom,
  author = {Fantom},
  year = 2026,
  title = {Fantom},
  url = {https://fantom.org/},
  lastaccessed = {2026-01-29}
}

@online{HomeProjectHaystack,
  author = {Project Haystack},
  year = 2026,
  title = {Project Haystack},
  url = {https://project-haystack.org/},
  lastaccessed = {2026-01-29}
}

@online{J2innHayson2025,
  author = {J2inn},
  year = 2026,
  title = {j2inn/hayson},
  url = {https://github.com/j2inn/hayson},
  lastaccessed = {2026-01-28}
}

@online{JsonProjectHaystack,
  author = {Project Haystack},
  year = 2026,
  title = {Json},
  url = {https://project-haystack.org/doc/docHaystack/Json\#v4},
  lastaccessed = {2026-01-28}
}

@online{LarkparserLark2026,
  author = {Lark},
  year = 2026,
  title = {Lark-Parser/Lark},
  url = {https://github.com/lark-parser/lark},
  lastaccessed = {2026-01-28}
}

@article{luoOverviewDataTools2021,
  title = {An Overview of Data Tools for Representing and Managing Building Information and Performance Data},
  author = {Luo, Na and Pritoni, Marco and Hong, Tianzhen},
  year = 2021,
  month = sep,
  journal = {Renewable and Sustainable Energy Reviews},
  volume = {147},
  pages = {111224},
  issn = {1364-0321},
  doi = {10.1016/j.rser.2021.111224},
  urldate = {2026-01-09}
}

@article{lygerakisKnowledgeGraphsOntologies2022,
  title = {Knowledge {{Graphs}}' {{Ontologies}} and {{Applications}} for {{Energy Efficiency}} in {{Buildings}}: {{A Review}}},
  shorttitle = {Knowledge {{Graphs}}' {{Ontologies}} and {{Applications}} for {{Energy Efficiency}} in {{Buildings}}},
  author = {Lygerakis, Filippos and Kampelis, Nikos and Kolokotsa, Dionysia},
  year = 2022,
  month = jan,
  journal = {Energies},
  volume = {15},
  number = {20},
  pages = {7520},
  publisher = {Multidisciplinary Digital Publishing Institute},
  issn = {1996-1073},
  doi = {10.3390/en15207520},
  urldate = {2026-01-09},
  copyright = {http://creativecommons.org/licenses/by/3.0/},
  langid = {english}
}

@online{NormalizationProjectHaystack,
  author = {Project Haystack},
  year = 2026,
  title = {Normalization},
  url = {https://project-haystack.org/doc/docHaystack/Normalization},
  lastaccessed = {2026-01-28}
}

@misc{paulsonComparisonBrickSchema2021,
  title = {A {{Comparison}} of the {{Brick Schema}} and {{Project Haystack}}},
  author = {Paulson, Erik},
  year = 2021,
  month = apr,
  journal = {Medium},
  urldate = {2024-11-05},
  langid = {english}
}

@article{pengBuildingOntologies45GDHC2025,
  title = {Building Ontologies for 4-{{5GDHC}}: {{A}} Critical Evaluation and Modeling Experiments on Building-Side Components},
  shorttitle = {Building Ontologies for 4-{{5GDHC}}},
  author = {Peng, Zeng and Ohlson Timoudas, Thomas and Wang, Qian},
  year = 2025,
  month = nov,
  journal = {Journal of Building Engineering},
  volume = {114},
  pages = {114204},
  issn = {2352-7102},
  doi = {10.1016/j.jobe.2025.114204},
  urldate = {2026-01-09}
}

@online{PointsProjectHaystack,
  author = {Project Haystack},
  year = 2026,
  title = {Points},
  url = {https://project-haystack.org/doc/docHaystack/Points},
  lastaccessed = {2026-01-29}
}

@article{pritoniMetadataSchemasOntologies2021,
  title = {Metadata {{Schemas}} and {{Ontologies}} for {{Building Energy Applications}}: {{A Critical Review}} and {{Use Case Analysis}}},
  shorttitle = {Metadata {{Schemas}} and {{Ontologies}} for {{Building Energy Applications}}},
  author = {Pritoni, Marco and Paine, Drew and Fierro, Gabriel and Mosiman, Cory and Poplawski, Michael and Saha, Avijit and Bender, Joel and Granderson, Jessica},
  year = 2021,
  month = jan,
  journal = {Energies},
  volume = {14},
  number = {7},
  pages = {2024},
  publisher = {Multidisciplinary Digital Publishing Institute},
  issn = {1996-1073},
  doi = {10.3390/en14072024},
  urldate = {2026-01-15},
  copyright = {http://creativecommons.org/licenses/by/3.0/},
  langid = {english}
}

@online{ProjectHaystack,
  author = {Project Haystack},
  title = {About -- {{Project Haystack}}},
  urldate = {2026-01-29},
  lastaccessed = {2026-01-29},
  howpublished = {https://www.project-haystack.org/about},
  url = {https://www.project-haystack.org/about},
  year = 2026
}

@online{ProjectHaystackHaystackdefs2025,
  author = {Project Haystack},
  title = {Haystack 4 Defs},
  year = 2025,
  month = dec,
  urldate = {2026-01-28},
  copyright = {AFL-3.0},
  howpublished = {Project Haystack},
  lastaccessed = {2026-01-28},
  url = {https://github.com/Project-Haystack/haystack-defs},
}

@article{qiangSystematicComparisonEvaluation2023,
  title = {A Systematic Comparison and Evaluation of Building Ontologies for Deploying Data-Driven Analytics in Smart Buildings},
  author = {Qiang, Zhangcheng and Hands, Stuart and Taylor, Kerry and Sethuvenkatraman, Subbu and Hugo, Daniel and Ghiasnezhad Omran, Pouya and Perera, Madhawa and Haller, Armin},
  year = 2023,
  month = aug,
  journal = {Energy and Buildings},
  volume = {292},
  pages = {113054},
  issn = {0378-7788},
  doi = {10.1016/j.enbuild.2023.113054},
  urldate = {2024-11-05}
}

@article{quinnCaseStudyComparing2021,
  title = {A Case Study Comparing the Completeness and Expressiveness of Two Industry Recognized Ontologies},
  author = {Quinn, Caroline and McArthur, J. J.},
  year = 2021,
  month = jan,
  journal = {Advanced Engineering Informatics},
  volume = {47},
  pages = {101233},
  issn = {1474-0346},
  doi = {10.1016/j.aei.2020.101233},
  urldate = {2026-01-09}
}

@misc{quinnComparisonBrickProject2022,
  title = {Comparison of {{Brick}} and {{Project Haystack}} to {{Support Smart Building Applications}}},
  author = {Quinn, Caroline and McArthur, J. J.},
  year = 2022,
  month = aug,
  number = {arXiv:2205.05521},
  eprint = {2205.05521},
  publisher = {arXiv},
  doi = {10.48550/arXiv.2205.05521},
  urldate = {2024-11-05},
  archiveprefix = {arXiv}
}

@inproceedings{SemanticsValidationRecursive,
  title={Semantics and validation of recursive SHACL},
  author={Corman, Julien and Reutter, Juan L and Savkovi{\'c}, Ognjen},
  booktitle={International Semantic Web Conference},
  pages={318--336},
  year={2018},
  organization={Springer}
}

@inproceedings{SHACLDescriptionLogic,
  title={SHACL: A description logic in disguise},
  author={Bogaerts, Bart and Jakubowski, Maxime and Van den Bussche, Jan},
  booktitle={International Conference on Logic Programming and Nonmonotonic Reasoning},
  pages={75--88},
  year={2022},
  organization={Springer}
}

@online{ShapesConstraintLanguage2017,
  author = {W3C},
  year = 2026,
  title = {Shapes {{Constraint Language}} ({{SHACL}})},
  url = {https://www.w3.org/TR/shacl/},
  lastaccessed = {2026-01-29}
}

@online{SystemsProjectHaystack,
  author = {Project Haystack},
  year = 2026,
  title = {Systems},
  url = {https://project-haystack.org/doc/docHaystack/Systems},
  lastaccessed = {2026-01-28}
}

@online{TrioProjectHaystack,
  author = {Project Haystack},
  year = 2026,
  title = {Trio},
  url = {https://project-haystack.org/doc/docHaystack/Trio},
  lastaccessed = {2026-01-28}
}

@online{ValidatorsPydanticValidation,
  author = {Pydantic},
  year = 2026,
  title = {Validators},
  url = {https://docs.pydantic.dev/latest/concepts/validators/},
  lastaccessed = {2026-01-29}
}

@online{WeatherProjectHaystack,
  author = {Project Haystack},
  year = 2026,
  title = {Weather},
  url = {https://project-haystack.org/doc/docHaystack/Weather},
  lastaccessed = {2026-01-29}
}

@online{WelcomePydanticPydantica,
  author = {Pydantic},
  year = 2026,
  title = {Pydantic},
  url = {https://docs.pydantic.dev/latest/},
  lastaccessed = {2026-01-29}
}

@article{yefiMetamEnThObjectOrientedMetamodel2024,
  title = {{{MetamEnTh}}: {{An Object-Oriented Metamodel}} for {{IoT Systems}} in {{Buildings}}},
  shorttitle = {{{MetamEnTh}}},
  author = {Yefi, Peter and Menon, Ramanunni Parakkal and Eicker, Ursula and Gu{\'e}h{\'e}neuc, Yann-Ga{\"e}l},
  year = 2024,
  month = aug,
  journal = {IEEE Internet of Things Journal},
  volume = {11},
  number = {15},
  pages = {25818--25838},
  issn = {2327-4662},
  doi = {10.1109/JIOT.2024.3373330},
  urldate = {2026-01-09}
}

@article{alfalouji2022iot,
  title={IoT middleware platforms for smart energy systems: an empirical expert survey},
  author={Alfalouji, Qamar and Schranz, Thomas and K{\"u}mpel, Alexander and Schraven, Markus and Storek, Thomas and Gross, Stephan and Monti, Antonello and M{\"u}ller, Dirk and Schweiger, Gerald},
  journal={Buildings},
  volume={12},
  number={5},
  pages={526},
  year={2022},
  publisher={MDPI}
}
